\documentclass[prl,twocolumn,showpacs,aps]{revtex4}

\usepackage{graphicx,color,textcomp}
\usepackage{amsmath}
\usepackage{psfrag,overpic,pst-all}
\usepackage[english,welsh]{babel}
\newcommand{\sfigt}[1]{$\,#1$}
\newcommand{\sfigf}[1]{$\textbf{#1}$}
\newcommand{\rmi}{{\mathrm i}}
\newcommand{\rme}{{\mathrm e}}
\newcommand{\pscpkm}{\ensuremath{\mathrm{ps^2\!/km}}}
\newcommand{\pscupkm}{\ensuremath{\mathrm{ps^3\!/km}}}

\begin{document}

\title{Impact of third order dispersion on nonlinear bifurcations in optical resonators}

\author{Fran\c{c}ois Leo}
\email{francois.leo@intec.ugent.be}
\affiliation{Service OPERA-\textit{photonique}, Universit\'e libre de  Bruxelles (U.L.B.), 50~Avenue F. D. Roosevelt, CP 194/5, B-1050 Bruxelles, Belgium}
\affiliation{ Photonics Research Group, Department of Information Technology, Ghent University-IMEC, Ghent B-9000, Belgium}

\author{St\'{e}phane Coen}
\affiliation{Department of Physics, The University of Auckland,  Private Bag, 92019, Auckland, New Zealand}       

\author{Pascal Kockaert}
\author{Philippe Emplit}
\author{Marc Haelterman}
\affiliation{Service OPERA-\textit{photonique}, Universit\'e libre de  Bruxelles (U.L.B.), 50~Avenue F. D. Roosevelt, CP 194/5, B-1050 Bruxelles, Belgium}

\author{Arnaud Mussot}
\author{Majid Taki}
\affiliation{PhLAM, Universit\'{e} de Lille~1,B\^{a}t.~P5-bis; UMR CNRS/USTL 8523, F-59655 Villeneuve d'Ascq, France}

\begin{abstract}
It is analytically shown that symmetry breaking, in dissipative systems,  affects the nature of the bifurcation at onset of instability resulting in transitions from super to subcritical bifurcations. In the case of a nonlinear fiber cavity, we have derived an amplitude equation to describe the nonlinear dynamics above threshold. An analytical expression of the critical transition curve is obtained and the predictions are in excellent agreement with the numerical solutions of the full dynamical model. 

\end{abstract}

\maketitle

Dissipative structures arise in many different fields of nonlinear science~\cite{Cross_Pattern_1993}. They are stable patterns that arise far from equibrium in dissipative systems. They exist because of the interplay between diffusion/diffraction and nonlinearity on one hand and between losses and an external source on the other hand. It was shown that such structures are significantly modified when a symmetry breaking is present in the system, i. e., odd-order spatial (for diffractive/diffusive systems) or temporal (for dispersive/temporal systems) derivatives leading to convective drift, or walk off terms depending on the specific physical situations. 
The effect of convective terms on dissipative structures has attracted a lot of interest in hydrodynamics~\cite{huerre_local_1990}, plasma physics~\cite{Briggs_electron_1964}, traffic flow~\cite{mittarai_spatiotemporal_2000} and nonlinear optics~\cite{izus_pattern_1999}. 
The main focus has been on the induced drift and the resulting convectively or absolutely unstable regimes~\cite{chomaz_absolute_1992}. More recently, there have been some studies on the nonlinear symmetry breaking induced by those terms in nonlinear optical dissipative systems~\cite{Zambrini_convection_2005,Leo_Nonlinear_2013}. They showed that albeit the linear nature of the broken symmetry, the nonlinearity of the system is affected, leading to spectral asymmetry and power dependent velocities of the traveling waves.

Here, we report on an important aspect of symmetry breaking that has not yet been investigated that drastically affects nonlinear bifurcations occurring in dissipative systems.
Nonlinear bifurcations have been intensely studied since the B\'enard experiment on the emergence of convection cells~\cite{Benard_Tourbillons_1901} and its theoretical study by lord Rayleigh~\cite{Rayleigh_Convection_1916}. Further theoretical developments, now commonly known as bifurcation theory~\cite{Nicolis_Self_1977,Manneville_Instabilities_1994}, have shown that through a  multi-scale approach, one can derive a simple ordinary differential equation (ODE) that describes the time evolution of the amplitude of the cells above threshold. This so-called amplitude equation reads~\cite{Cross_Pattern_1993,Manneville_Instabilities_1994}~: 
\begin{equation}\label{EqBenard}
\frac{dA}{dt} = rA-sA^3,
\end{equation}
where $A$ is the amplitude of the cell, $r$ is the relative distance to threshold and $s$ represents the nonlinear saturation. Note that the absence of even order terms is due to the symmetry $A\,\rightarrow\,-A$ of the system. The trivial solution $A=0$ is unstable for $r>0$ and the system evolves towards cells with an amplitude $A=\pm\sqrt{r/s}$, aligned in a pitchfork configuration.
From this we immediately see the critical role played by the sign of $s$. In the case $s<0$, the nonlinear term is self limiting. The solutions exist above threshold for $r>0$ and the bifurcation is called supercritical. In the case $s>0$, the non-linear term is self amplifying. The solutions exist for $r<0$ and the bifurcation is called subcritical. In this case, the dissipative structures corresponding to $r>0$ can be calculated by adding higher order terms, providing nonlinear saturation, on the right-hand side of Eq.~\ref{EqBenard}. This leads to a bistable regime between patterned and unpatterned solutions that can lead to the formation of localized structures.
Here we perform, to the best of our knowkedge, the first analytical study of the effect of convective terms on the sign of the nonlinear term $s$ in the case of modulated patterns appearing in optical resonators. 
We consider a passive nonlinear resonator whose dynamics is well described, in the mean field approximation, by the well known Lugiato-Lefever (LL) equation~\cite{lugiato_spatial_1987}. 
In a recent study, we analytically and experimentally investigated the spectral asymmetry induced by a convective term~\cite{Leo_Nonlinear_2013}.
We showed that the third-order dispersion induced an asymmetry in the intensities of high-order harmonics resulting in a transition from symmetric to asymmetric dissipative structures. In this work, we show how the symmetry breaking actually modifies the intrinsic nonlinearity of the system. More precisely, we show that it affects the bifurcation nature, at onset of the instability, leading to \textit{a transition from a sub to a supercritical bifurcation} which drastically impacts the subsequent dynamics above threshold. The analytical description of this bifurcation transition, based on the amplitude equation of the passive optical cavity, demonstrates an original dependence of the nonlinear saturation term $s$ upon the symmetry breaking term (here the third-order dispersion term). 

We start from the dimensionless mean-field equation describing
nonlinear resonators near the zero-dispersion wavelength~\cite{Pedro_Third_2014}~:
\begin{multline}\label{eqLLadim}
  \partial_t E(t,\tau)=\left[-1 +{\rm{i}}(|E(t,\tau)|^2-\Delta)-{\rm{i}}\eta\partial^2_{\tau}\right.\\
  \left.+d_3\partial^3_{\tau}\right]E(t,\tau)+S,
\end{multline}
where $S$ and $E$ are the normalized slowly varying
envelopes for pump and signal fields respectively, $\Delta$ is the
normalized cavity detuning, $\eta$ is the sign of the second-order dispersion
term (SOD) and $d_3$ is the normalized third-order dispersion coefficient. Details on the normalization can be found in \cite{Leo_Temporal_2010,Pedro_Third_2014}.
Both convective and absolute instabilities have been recently reported
for the steady-state solution $E_s$ satisfying the equation
$S=\left[1+i(\Delta-I_s)\right]E_s$ where $I_s=|E_s|^2=E_s^2$ ($E_s$ is taken to be real)
\cite{mussot_optical_2008}. In this paper, we devote our attention to the monostable case (i.e. $\Delta<\sqrt{3}$) with anomalous dispersion where Eq.~(\ref{eqLLadim}) exhibits a bifurcation at $I_s = 1$, corresponding to a pump power $X_s = |S|^2 = 2 - 2\Delta + \Delta^2$.
The system then evolves towards modulated solutions characterized by a frequency
$\Omega_c =\sqrt{(\Delta-2)/\eta}$ and a wave vector ${\kappa}_c = -
d_3\Omega_c^3$ \cite{mussot_optical_2008}.
When $d_3=0$, Eq.~(\ref{eqLLadim}) reduces to the well-know LL equation describing nonlinear spatial cavities \cite{lugiato_spatial_1987}. It has been shown that the modulated stationary solutions of that equation can be analytically calculated by the standard method of bifurcation theory~\cite{Nicolis_Self_1977,Manneville_Instabilities_1994}. From this study comes the well-known transitional detuning $\Delta_t=41/30$, characterizing the passing from a supercritical bifurcation, where the modulated solutions appear above threshold for pump powers higher than $X_s$ and are stable, to a subcritical bifurcation where the modulated solutions appear above threshold for pump powers lower than $X_s$ and are unstable.

\begin{figure}[figureplacement=h]
\psfrag{a}[][]{\sfigf{a}}
\psfrag{b}[][]{\sfigf{b}}
\psfrag{c}[][]{\sfigf{c}}
\psfrag{d}[][]{\sfigf{d}}
\psfrag{X}[t][b][1.2]{$X$} 
\psfrag{P}[c][c][1.2][-90]{$|D|$}
\psfrag{Q}[c][c][1.2][-90]{$|A^{\pm}|\:\:\:\:\:\:\:\:$}
\psfrag{W}[c][c][1.2][-90]{$|C^{\pm}|\:\:\:\:\:\:$}
\psfrag{V}[c][c][1.2][-90]{$v\:\:\:$}
\centering
\begin{overpic}[grid = false,clip]{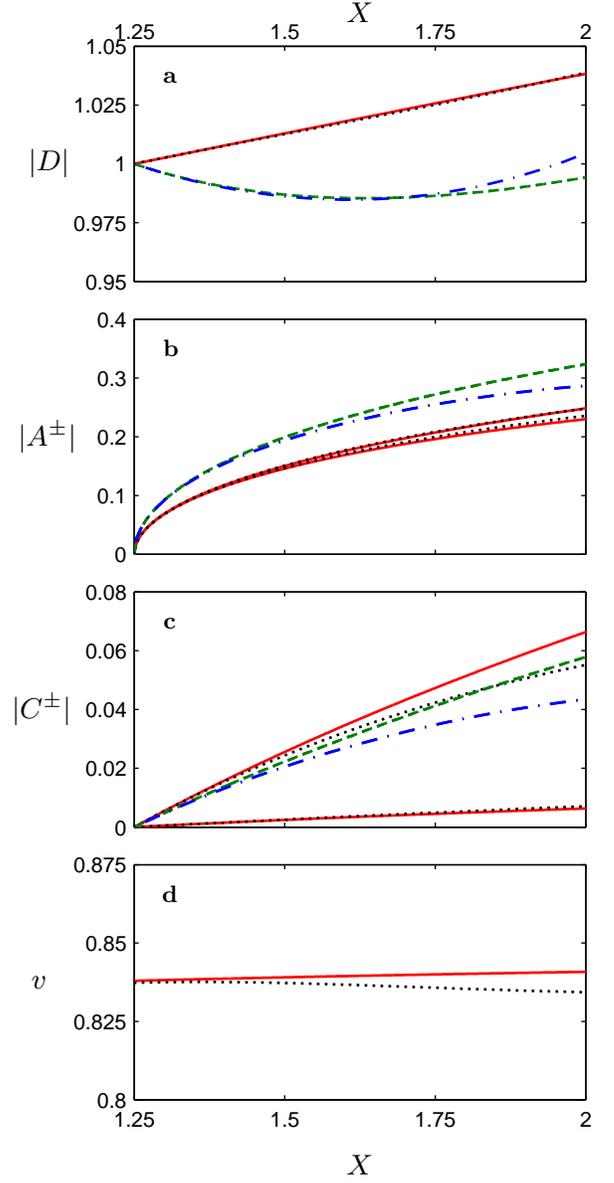}
\end{overpic}
\renewcommand{\figurename}{Fig.}
\caption{\sfigf{a}\,-\sfigf{c}, Spectral amplitudes of the dissipative structures for a cavity with $\Delta =0.5$ and $d_3 = 0.56$ (solid line: analytical results and dotted line: numerical results) or $d_3 = 0$ (dashed line: analytical results and dashed-dotted line : numerical results). \sfigf{d}, Speed of the traveling structures for $d_3 = 0.56$. Note that the dimensionless $d_3$ parameter is extracted from the following physical parameters corresponding to a recently reported experimental setup : $\theta=0.1$, $\alpha_fL=0.22$, $L=105$\,m, $\alpha = 0.16$, $\Delta =0.5$, $\beta_2 = -0.25\,\pscpkm$ and $\beta_3 = 0.12\,\pscupkm$~\cite{Leo_Nonlinear_2013}.}
\label{figasym}
\end{figure}

In order to study the effect of the third-order dispersion
on the stationary modulated solutions and the transitional detuning, we perform the same multi-scale analysis
right above the instability threshold but now including third-order dispersion.
We start by expanding the variables in multiple orders of a small parameter $\varepsilon$, defined as $\varepsilon^2 = I_s - 1$, that is the distance from the instability threshold. The envelope of the electric field is rewritten in terms of the amplitudes \(a_k\), defined by $E=E_s+\varepsilon a_1+ \varepsilon^{2}a_2+\varepsilon ^{3}a_3+...$.  Following the approach of~\cite{Nicolis_Self_1977,Manneville_Instabilities_1994}, and taking into account the gain spectrum of the instability~\cite{haelterman_dissipative_1992} we expand the fast time $t$ and the slow time $\tau$. We introduce the new times : $T_0 =t$, $T_1=\varepsilon t$, $T_2=\varepsilon^2 t$, $\tau_0 =\tau$ and $\tau_1=\varepsilon \tau$ so that the corresponding temporal derivatives become $\partial_t = \partial_{T_0}+\varepsilon\partial_{T_1}+\varepsilon^2\partial_{T_2}$ and $\partial_\tau = \partial_{\tau_0}+\varepsilon\partial_{\tau_1}$. We then assume that the amplitudes \(a_k\) is the sum of quasi-monochromatic waves written in the form $a_1 = \left( A_1{\rme}^{{\rmi}\left({\Omega_c}\tau_0+{\kappa_c}T_0\right)}+ A_1^*{\rme}^{-{\rmi}\left({\Omega_c}\tau_0+{\kappa_c}T_0\right)}\right)$ where $A_{1}$ and its complex conjugate $A_{1}^{*}$ are slowly varying amplitudes, and $a_k = D_k + A_k^+{\rme}^{{\rmi}\left({\Omega_c}\tau_0+{\kappa_c}T_0\right)}+ A_k^-{\rme}^{-{\rmi}\left({\Omega_c}\tau_0+{\kappa_c}T_0\right)}+ C_k^+{\rme}^{2{\rmi}\left({\Omega_c}\tau_0+{\kappa_c}T_0\right)}+
C_k^-{\rme}^{-2{\rmi}\left({\Omega_c}\tau_0+{\kappa_c}T_0\right)}$ with k = 2,3.
This form of \(a_1\) is justified by the fact that right above the instability threshold, the gain exceeds unity only in the vicinity of \(\Omega\approx\pm\,\Omega_c\), while for the second- and third-order corrections, contributions at \(0,\pm\,2\Omega_c\) appear because of the nonlinear interactions.  By substitution of the above expansions in Eq.(\ref{eqLLadim}), we obtain a hierarchy of equations for the successive orders of $\varepsilon$. 
The evolution of $A_{1}$ is described by the following equation, obtained as a consequence of a solvability condition ~\cite{Nicolis_Self_1977,Manneville_Instabilities_1994} at the third order~:
\begin{align}
\partial_{t} A+3d_{3}\Omega_{c}^{2}\partial_{\tau} A=& (I_{s}-1)A+(2\Omega_{c}^{2}+3\rmi d_{3}\Omega_{c})\partial^{2}_{\tau}A \notag \\&+(s_{1}+\rmi s_{2})\vert A\vert^{2}A,
\label{Equation amplitude}
\end{align} 
\noindent where we have set $A=\varepsilon A_{1}$ and the parameters  are defined as
\begin{align}
s_{1}&=24\frac{2G+3}{G^{2}}+4\frac{G^{2}(1-2G)+H^{2}(2G-3)}{(G^{2}-H^{2})+4H^{2}} \label{d1},\\
s_{2}&=\frac{4H[2(1-2G)+G^{2}-H^{2}]}{(G^{2}-H^{2})+4H^{2}},
\end{align}
with
\begin{align}
G&=3(\Delta-2) \nonumber \\
H&=-6d_{3}\Omega_{c}^{3} \nonumber
\end{align}
This equation of complex Ginzburg-Landau type describes the time evolution of the Stokes wave (fundamental mode) above threshold. First, note that, in absence of TOD ($\beta_3=0 $), three terms in Eq.~(\ref{Equation amplitude}) disappear since $d_3=0$ and $s_{2}=0$. 
As a result, the presence of TOD drastically impacts the dynamics by introducing drift and diffraction effects (terms in Eq.~(\ref{Equation amplitude}) with $d_3 $) together with a nonlinear frequency modulation (term with $s_2$). More importantly, the presence of TOD affects the nature of the bifurcation as can be seen from Eq.~(\ref {d1}) since the nonlinear coefficient $s_{1}$ can change sign as it will be shown below. This clearly demonstrates how the symmetry breaking introduces a rich and a complex dynamics. However, in what follows, we only investigate the important dynamics resulting from transitions between super- and sub-critical bifurcations. Before proceeding further, and to give a complete analytical description above threshold, let us resolve the system up to the third order. After lengthy but straightforward calculations, we find the following analytical expression for the dissipative structures~: 
 \begin{multline}
E(t,\tau) =  D + A^{+}{\rme}^{{\rmi}\left({\Omega_c}\tau+{(\kappa_c+\kappa)}t\right)} + A^{-}{\rme}^{-{\rmi}\left({\Omega_c}\tau+{(\kappa_c+\kappa)}t\right)}\\
+ C^{+}{\rme}^{2{\rmi}\left({\Omega_c}\tau+{(\kappa_c+\kappa)}t\right)} + C^{-}{\rme}^{-2{\rmi}\left({\Omega_c}\tau+{(\kappa_c+\kappa)}t\right)},
\label{Eqsolstat}
\end{multline} 
with
\begin{eqnarray}
D&=& E_s +\frac{12|A_{st}|^2}{G^2}\left[(2G+3) + 3{\rmi}(G+1)\right],\label{eqD}\\
A^+&=& (1+{\rmi})|A_{st}|\left(1-MH+{\rmi}N\right),\label{eqA+}\\
A^-&=& (1+{\rmi})|A_{st}|\left(1+MH+{\rmi}N\right),\label{eqA-}\\
C^+&=&2|A_{st}|^2\frac{1+{\rmi}-(2+{\rmi})(G-H)}{G^2-H^2-2{\rmi}H},\\
C^-&=&2|A_{st}|^2\frac{1+{\rmi}-(2+{\rmi})(G+H)}{G^2-H^2+2{\rmi}H},\\
{\kappa} &=& s_{2}|A_{st}|^2,\\
|A_{st}|^2 &=& (1-I_s)]/s_{1}\label{eqSst},\\
M&=&|A_{st}|^2\frac{12H-20GH}{(G^2-H^2)^2+4H^2},\\
N&=&Is-1+\nonumber\\
&& |A_{st}|^2\left(3+12\frac{7G+9}{G^2}
+ \frac{(6-10G)(G^2-H^2)}{(G^2-H^2)^2+4H^2}\right),\nonumber\\
\end{eqnarray}
where $|A_{st}|$ and $\kappa$ are the amplitude and the nonlinear correction to the wavevector of the nonlinear dissipative solution of Eq. (\ref{Equation amplitude}).
Here we purposely neglected the dependence of the spectral amplitudes in the fast time $\tau$ as we are only interested in the leading contribution of each harmonics.
From Eq.~(\ref{Eqsolstat}), we can deduce the speed of the travelling waves. We get~: 
\begin{equation}
v = -\frac{\kappa_c+\kappa}{\Omega_c} = d_3\Omega_c^2 + \frac{s_{2}}{\Omega_cs_{1}}(I_s-1)
\end{equation}
Let us note that there is a correction term to the usual convection velocity (the first term on the right-hand side) depending linearly on the distance to the threshold as previously pointed out in the case of nonlinear optical parametric oscillators~\cite{Zambrini_convection_2005}.
The five main spectral modes and the speed of the inhomogeneous solutions are represented on Fig.~\ref{figasym} and compared to numerical resolutions of Eq.~(\ref{eqLLadim}). 
As can be seen, the agreement is very good up to almost twice the threshold. The values of the parameters are set to the ones of a recently experimentally implemented fiber cavity~\cite{Leo_Nonlinear_2013}.
\begin{figure}[figureplacement=h]
\psfrag{a}[][]{\sfigf{a}}
\psfrag{b}[][]{\sfigf{b}}
\psfrag{c}[][]{\sfigf{c}}
\psfrag{D}[t][b][1.2]{$\Delta$} 
\psfrag{X}[t][b][1.2]{$X$} \psfrag{Y}[c][c][1.2][-90]{$s_{1}\:\:\:$}
\psfrag{Z}[t][b][1.2]{$d_3$} \psfrag{W}[c][c][1.2][-90]{$\Delta_t\:\:$}
\psfrag{P}[t][b][1.2][-90]{$P_{max}\:\:\:\:\:$}
\centering
\begin{overpic}[grid = false,clip]{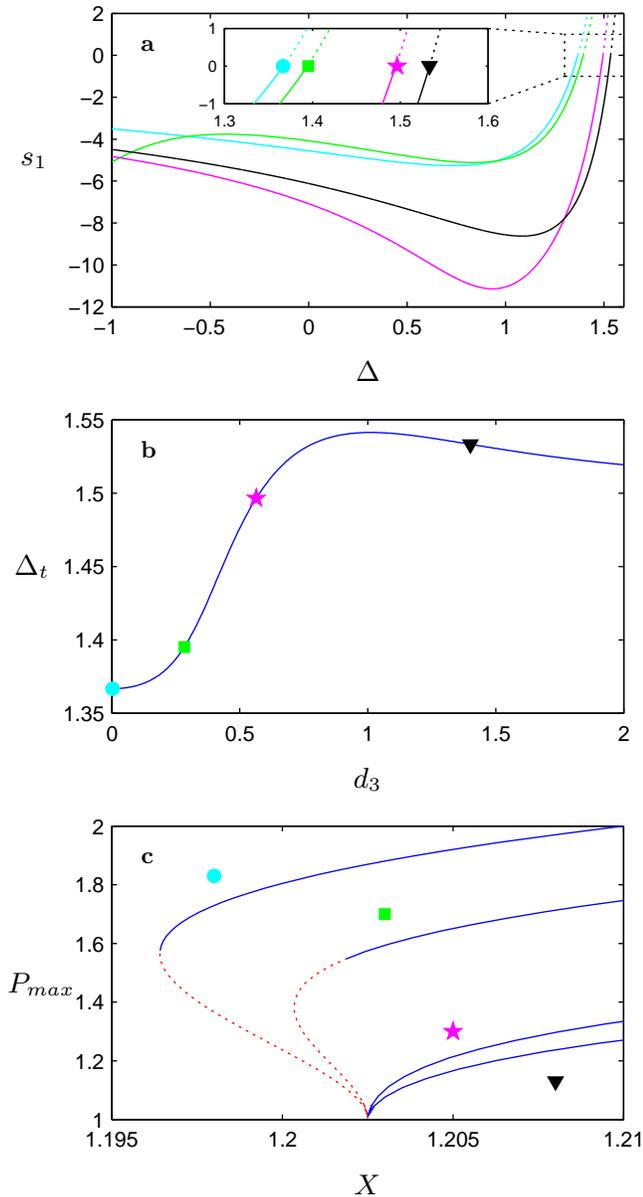}
\end{overpic}
\renewcommand{\figurename}{Fig.}
\caption{Study of the nature change of the bifurcations. \sfigf{a}, Evolution of $s_{1}$ with the detuning for different TOD. $d_3 = 0$~(circle), $d_3 = 0.28 $ (square), $d_3 = 0.56$ (star) and $d_3 = 1.4$ (triangle). The change of sign of $s_{1}$ means a change of bifurcation nature. \sfigf{b} Transitional detuning deduced from the change of sign of $s_{1}$. \sfigf{c} Numerical computation of the maximum power of the dissipative structures with and without TOD, for a detuning $\Delta = 1.45$, showing different types of bifurcations. The continuous blue ( resp. dotted red) line corresponds to stable (resp. unstable) solutions of Eq.~(\ref{eqLLadim}).}
\label{figbif}
\end{figure}
 
We clearly observe, from the figures, the asymmetry induced by the TOD on both the first and second harmonics. As recently demonstrated experimentally, we find that the asymmetry is much stronger on the second harmonics~\cite{Leo_Nonlinear_2013}.
Moreover, these results show an important difference in the average power of the harmonics when the TOD is taken into account. This means that the TOD is not only responsible for the drift and the symmetry breaking but also for important changes in the modulation. Furthermore, we see from Eqs.~(\ref{eqD}-\ref{eqSst}) and Eq.~(\ref{d1}) that there are inhomogeneous solutions appearing above threshold for pump powers higher (resp. lower) than $X_s$ if $s_{1}$ is negative (resp. positive). As stated before, the bifurcation is then supercritical (resp. subcritical) and those solutions are stable (resp. unstable). In the case $d_3=0$, Eq.~(\ref{d1}) reduces to 
$s_{1}=4\left(30\Delta-41\right)/(3(\Delta-2))^2$ so that the bifurcation becomes subcritical when the detuning is higher than the transitional detuning $\Delta_t = 41/30$ \cite{lugiato_spatial_1987}. When $d_3\neq0$ however, the transition [relation resulting from $s_{1}=0$ in Eq. (\ref{d1})] does not occur at a fixed value as the critical detuning depends on the TOD. The evolution of the parameter $s_{1}$ with the detuning for different values of the TOD is represented on Fig.~\ref{figbif}\sfigt{a}. 
The detuning $\Delta_t$ corresponding to the change in sign of $s_{1}$ is represented on Fig.~\ref{figbif}\sfigt{b} where we observe an increase of the region corresponding to a supercritical bifurcation as the TOD increases. To confirm this behaviour, we performed a numerical computation of the nonlinear solutions for $\Delta = 1.45$ using a Newton-Raphson solver.
The bifurcation curves representing the peak power of the modulated solutions above threshold with and without the TOD are depicted on Fig.~\ref{figbif}\sfigt{c}. This result is in excellent agreement with analytical predictions stating that the nature of the bifurcation depends on the third-order dispersion. Note that this behaviour is reminiscent of the inhibition of the bistablity reported in previous studies of convective instabilities~\cite{chomaz_absolute_1992,Coen_Convection_1999,Pedro_Third_2014}.
Moreover, this inhibition of bistability between the patterned and continous solution is expected to have an impact on the formation of temporal solitons~\cite{Leo_Temporal_2010} in optical resonators~\cite{mussot_optical_2008} as such localized dissipative structures are supported by this bistable regime~\cite{Scroggie_Pattern_1994}. Furthermore, it was recently shown that microresonator based frequency combs are linked to such localized structures~\cite{Coen_Modeling_2013}. As the TOD is critical in the case of octave spanning frequency combs, Eq.~(\ref{eqLLadim}) is commonly used to simulated frequency comb generation. Very recent studies have already shown that the TOD impacts the bifurcation structure of frequency combs~\cite{Pedro_Third_2014,Milian_Soliton_2014}. We believe that our results will trigger further studies on the impact of TOD on microresonator based Kerr frequency combs.
In summary, our analysis establishes the important impact of the symmetry breaking on bifurcation that arises at onset of the instability in a passive optical cavity. This symmetry breaking is induced by the third-order dispersion, resulting in a drastic change in the above threshold dynamics. We have demonstrated that depending on the values of the third-order dispersion, transitions from sub to supercritical bifurcations can occur.  
We have obtained an exact analytical expression of the critical transition curve and derived an amplitude equation to describe the nonlinear dynamics above threshold. Our analytical predictions are in excellent agreement the numerical simulations of the Lugiato-Lefever equation. We believe that the results here are not specific to optical cavities but can be applied to the large class of nonlinear dynamical systems presenting reflection (or time-reversal) symmetry breaking.

\end{document}